\documentclass{article}
\usepackage{spconf,amsmath,graphicx}

\usepackage{enumitem}
\usepackage{lipsum} 
\setlist{nosep, leftmargin=14pt}
\usepackage{xcolor}
\usepackage{hyperref}

\usepackage{mwe} 

\title{Interpretable Embeddings for Segmentation-Free Single-Cell Analysis in Multiplex Imaging}
\name{%
\begin{tabular}{@{}c@{}}
Simon Gutwein$^{1,2,3,4,5,6}$ \qquad 
Daria Lazic$^{1}$ \qquad 
Thomas Walter$^{4,5,6}$ \\ 
Sabine Taschner-Mandl$^{1}$ \qquad 
Roxane Licandro$^{3}$ \qquad 
\end{tabular}
}

\address{
    $^{1}$St. Anna Children's Cancer Research Institute, Vienna, Austria \\
    $^{2}$TU Wien, Institute of Visual Computing and Human-Centered Technology, CVL, Vienna, Austria \\
    $^{3}$Medical University of Vienna, Biomedical Imaging and Image-guided Therapy, CIR, Vienna, Austria \\
    $^{4}$Centre for Computational Biology (CBIO), Mines Paris, PSL University, 75006 Paris, France \\
    $^{5}$Institut Curie, 75248 Paris Cedex, France \\
    $^{6}$INSERM, U900, 75248 Paris Cedex, France
}

\begin{document}
\maketitle
\begin{abstract}
Multiplex Imaging (MI) enables the simultaneous visualization of multiple biological markers in separate imaging channels at subcellular resolution, providing valuable insights into cell-type heterogeneity and spatial organization. However, current computational pipelines rely on cell segmentation algorithms, which require laborious fine-tuning and can introduce downstream errors due to inaccurate single-cell representations. We propose a segmentation-free deep learning approach that leverages grouped convolutions to learn interpretable embedded features from each imaging channel, enabling robust cell-type identification without manual feature selection. Validated on an Imaging Mass Cytometry dataset of 1.8 million cells from neuroblastoma patients, our method enables the accurate identification of known cell types, showcasing its scalability and suitability for high-dimensional MI data.
\end{abstract}

\section{Introduction}
\label{sec:intro}
Multiplex Imaging (MI) has gained significant attention in recent years with the introduction of techniques like CODEX~\cite{Black2021}, MIBI~\cite{Ptacek2020}, and Imaging Mass Cytometry (IMC)~\cite{Giesen2014}, which enable the visualization of dozens of protein markers, each displayed in its own individual image channel. However, MI presents new challenges due to its high dimensionality. In MI, single-cell characterization is driven by expression profiles of specific surface proteins, rather than by morphology or localization patterns.\\
Quantifying and classifying individual cells from image data has a long history, with numerous methods developed for subcellular protein localization \cite{Boland1998} and morphological phenotype classification \cite{Carpenter2006,Bakal2007}. Despite the introduction of IMC almost a decade ago, methods for analyzing MI images still depend heavily on cell segmentation, using manually crafted features such as mean intensity per image channel over segmentation masks as a proxy for protein expression \cite{Windhager2023,Thirumal2022,Bortolomeazzi2022,Hickey2021}. These manually crafted features remain common, because they allow experts to easily interpret protein expression using their domain knowledge for specific cell types.\\
However, these techniques require accurate segmentation, which is challenging with IMC images due to low spatial resolution, ambiguous cell boundaries, and high cell density. Common errors include merging multiple cells, splitting individual cells, and inaccurate border detection, leading to signal contamination from neighboring cells, especially in dense regions, thereby compromising downstream analyses. To mitigate some of these problems, considerable efforts have been made to optimize segmentation algorithms and reduce error rates, but this does not fully solve the issue.\\
To address these challenges, we propose a segmentation-free deep learning model with the following contributions:\\


\noindent \textbf{Segmentation-Free Analysis}: We bypass segmentation, using cell-centered patches to learn representations directly from full image content, avoiding manual feature extraction.\\

\noindent \textbf{Interpretability-Focused Design}: Our proposed architecture enables experts to directly interpret the model’s outputs by tracking the influence of individual image channels on the embedding space, offering crucial insights otherwise hidden by the complexity of MI data.\\

\noindent \textbf{Validation on Real-World Neuroblastoma IMC Dataset}: We validate our method’s applicability for cell phenotyping using an IMC dataset of 1.8 million cells collected from the bone marrow of neuroblastoma patients.\\
\begin{figure*}[!t]
    \centering
    \includegraphics[width=\textwidth]{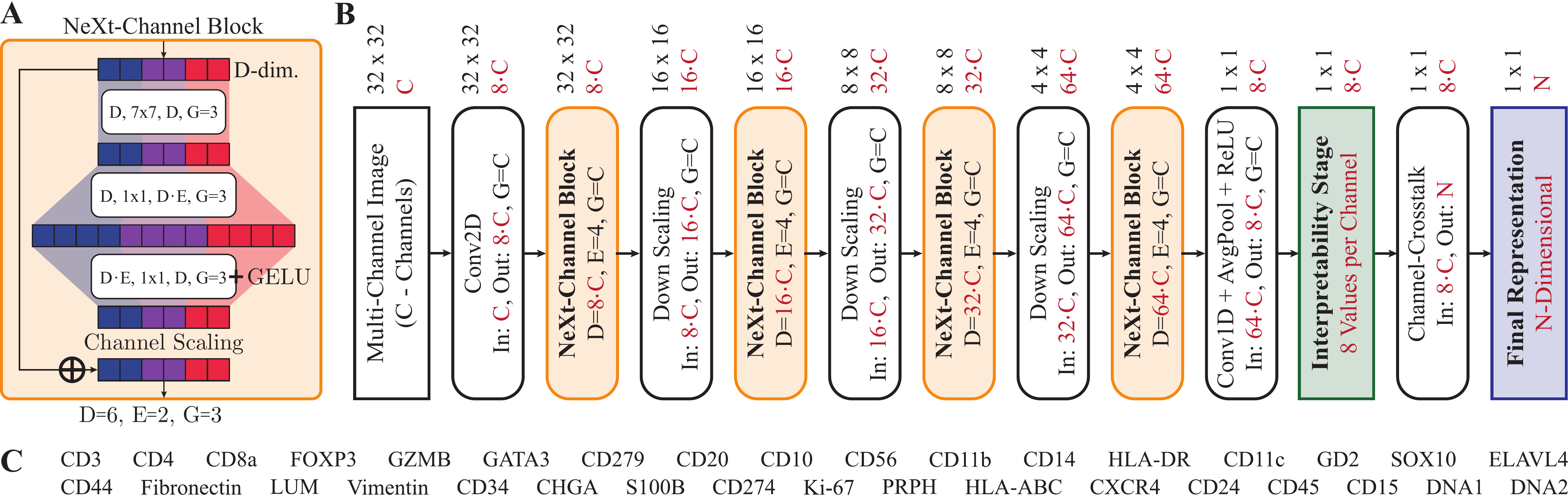}  
    \caption{\textbf{A}: Core unit of the proposed architecture. White blocks represent convolutional operations, formatted as: input features, $\text{kernel size} \times \text{kernel size}$, and output features. Here, $D$ is the number of input features, $G$ denotes the number of groups, and $E$ indicates the expansion factor of the block. \textbf{B}: Example data flow through the model, starting from a $32 \times 32 \times C$ pixel input patch, where $C$ is the number of image channels. White blocks represent convolutional layers with a $3 \times 3$ kernel size and number of groups ($G$). The interpretability stage (green) is used to obtain channel-specific activation, enabling expert interpretation. The final representation is generated by processing the previously disentangled channels together, allowing for crosstalk between them. \textbf{C}: Markers used in the IMC panel.}    
    \label{fig:model}
\end{figure*}
\newpage
\section{Methods}
\label{sec:methods}

To address the limitations of segmentation based approaches, we propose a deep learning model that uses cell-centered patches as input to learn human-interpretable representations.
Most Computer Vision tasks involve single-channel or three-channel images, where the information in each channel is highly correlated. In MI images, however, the marker associated channels are often weakly or not correlated. Consequently, conventional convolutions, which process all image channels simultaneously, result in an entanglement that obscures the contribution of individual channels to the embedding, limiting interpretability.\\ 

\noindent\textbf{Model Architecture:} To address this, we present the core component of our architecture, the \textit{NeXt-Channel} Block (Fig.~\ref{fig:model}~A), inspired by the ConvNeXt~\cite{Liu2022} structure. It leverages the unique properties of grouped convolutions to maintain channel-specific information disentangled. This design ensures that each biological marker is processed independently, enhancing interpretability and allowing domain experts to evaluate the contribution of individual markers to the cell representations. The NeXt-Channel Block aligns the number of grouped convolutions with the number of MI image channels. Each group processes its specific channel independently, preventing crosstalk and preserving the unique contribution of each marker. Fig.~\ref{fig:model}~A illustrates this concept with a 6-dimensional input, 3 groups, and an expansion factor (E) of 2, where the color coding indicates the strict separation of groups. For a detailed explanation of the NeXt-Channel block configuration, please refer to the ConvNeXt paper \cite{Liu2022}, which provides comprehensive information on the block design.\\
As shown in Fig.~\ref{fig:model}~B, the number of groups (G) is consistent across all layers, ensuring complete disentanglement of input channels (C).
This separation is maintained up to the "interpretability stage" (green in Fig.~\ref{fig:model}~B), where the feature space is structured such that a set of features is only influenced by its respective input channel. This design allows users to directly assess the contribution of each channel to specific features in the feature space. To enable interaction between different channels, the final layer integrates these interpretable features, allowing their information to be combined. This controlled entanglement facilitates cross-channel communication, producing a unified representation suitable for tasks such as clustering.\\

\noindent\textbf{IMC Dataset:} We evaluated our approach using an IMC dataset consisting of 674 images from 84 neuroblastoma patients, each containing 34 channels representing distinct biological markers (see full list in Fig.~\ref{fig:model}~C). Preprocessing involved spillover correction \cite{Chevrier2018} and background removal using ilastik \cite{Berg2019}. For model training, Cellpose \cite{Stringer2021} was employed to detect nuclei and extract $32 \times 32$ pixel patches centered around each detected cell, resulting in a total of 1.8 million single-cell patches.\\

\noindent\textbf{Training Procedure:} We trained the model using SimCLR \cite{Chen2020} with contrastive learning, applying five augmentation techniques to each patch: (1) intensity scaling (50\% to 200\%), (2) Gaussian noise (mean 0, std 0 to 0.1), (3) random rotations (0° to 360°), (4) random scaling (0.9 to 1.1) and (5) horizontal/vertical flips. Different receptive fields were simulated by extracting crops of 16, 16, 14, and 12 pixels from the center of each augmented patch. For each batch, 768 patches were sampled, generating 3072 views per batch. We used the LARS optimizer \cite{You2017} with a learning rate of 4.6, momentum of 0.9, and weight decay of $10^{-6}$. Training was conducted over 5000 epochs with cosine annealing after 10 warm-up epochs, yielding a total of 15.36 million views.\\

\noindent\textbf{Experimental Setup:} Due to the absence of ground truth labels in MI, we compared our method to a state-of-the-art benchmark analysis as outlined in \cite{Windhager2023}. The benchmark utilized a manually fine-tuned Cellpose~\cite{Stringer2021} segmentation on high-resolution nuclei images of the same slide, which is typically unavailable, followed by manual feature extraction (mean intensity per channel over the segmentation mask). In contrast, our method bypasses segmentation entirely, generating representations directly from raw input patches.\\
The model was configured with G=34 to match the number of image channels, an expansion factor of E=2, and an embedding dimension N=256 as shown in Fig.~\ref{fig:model}~B. Our cell phenotyping workflow involved: (1) learning cell representations from single-cell patches, (2) clustering the 256-dimensional embeddings using Phenograph~\cite{Levine2015} with \texttt{k=8} for graph construction, labeling clusters with fewer than 50 cells as unknown (class:~\texttt{-1}), (3) aggregating mean feature values per channel from the interpretability stage (see Fig.~\ref{fig:model}B, green layer), and (4) assigning phenotypes to predefined categories such as Monocytes/Dendritic Cells/Natural Killer Cells (MO/DC/NK), tumor cells, B cells, T cells, progenitor cells, granulocytes, and others. These categories were defined based on the benchmark analysis. Two domain experts performed phenotyping, following standard procedures. Finally, a confusion matrix was constructed to compare the rediscovery rates of phenotypes identified by our method against the benchmark segmentation-based approach.

\section{Results}
\label{sec:results}

\begin{figure}
    \centering
    \includegraphics[width=\linewidth]{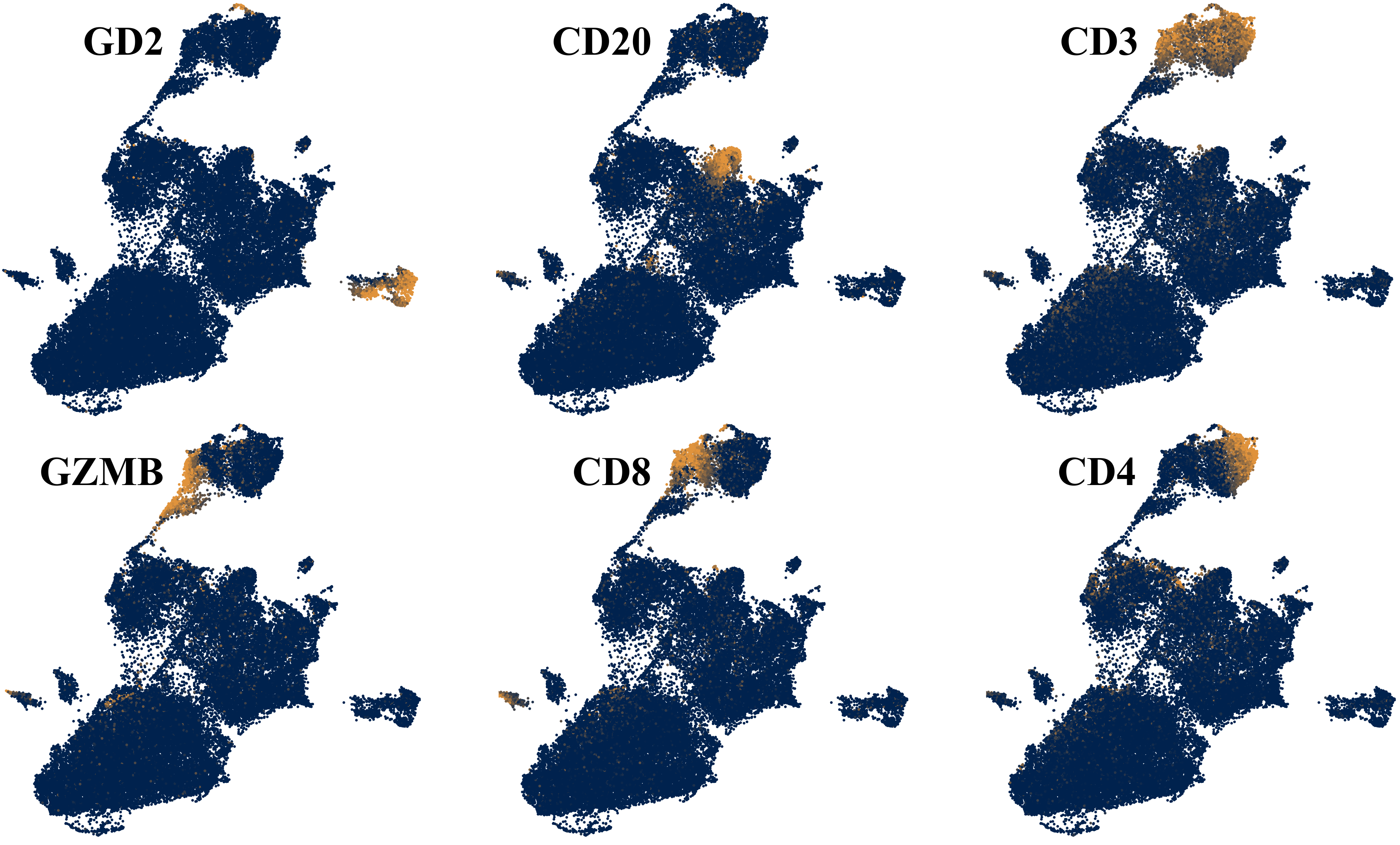}  
    \caption{UMAP plots showing the contribution of specific markers to the representation space for selected channels: CD20 (expressed on B cells), GD2 (expressed on neuroblastoma cells), CD3 (expressed on T cells), CD8 (expressed on cytotoxic T cells), CD4 (expressed on helper T cells) and GZMB (indicative of cytotoxic T cells and natural killer cells).}
    \label{fig:activation}
\end{figure}

\noindent\textbf{Qualitative and Quantitative Assessment of Representation Interpretability}: Our model effectively groups biologically similar cells in the representation space, as demonstrated by the marker-specific contributions in the UMAP~\cite{McInnes2020} visualization of 100,000 randomly sampled cells (Fig.~\ref{fig:activation}). Each UMAP plot displays the aggregation of patches with high contributions from specific markers, such as CD3, indicating a population of CD3+ (expressing CD3) cells.

\begin{figure}
    \centering
    \includegraphics[width=\linewidth]{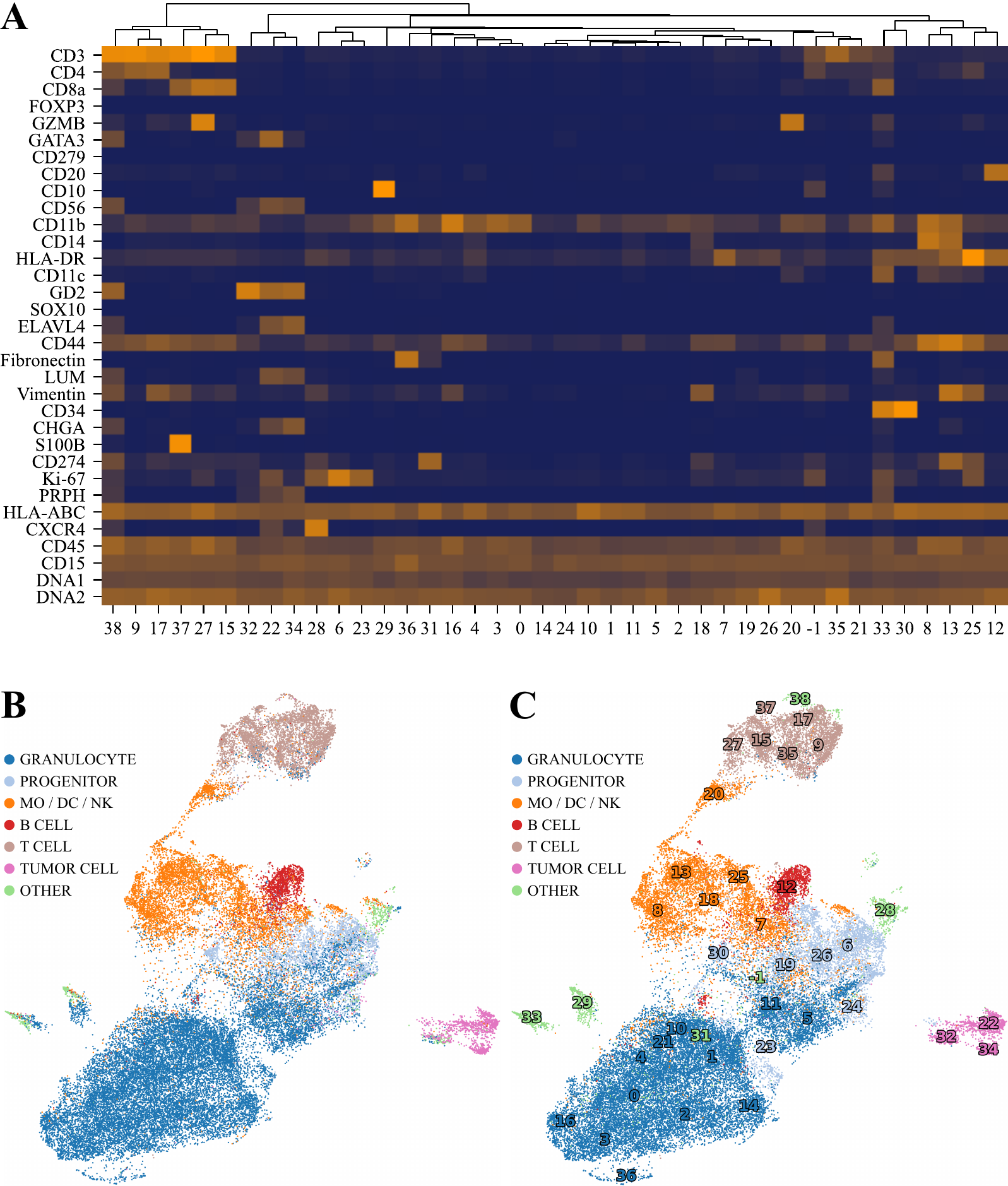}  
    \caption{\textbf{A}: Heatmap of average activations from the interpretability stage of our proposed method for each Phenograph cluster across all 34 markers in the panel. \textbf{B}: UMAP embedding ofthe representation stage of our method, colored by cell type based on benchmark phenotyping using nucleus segmentation. \textbf{C}: Same UMAP as in (B), but with Phenograph clusters and their cell type assignments as identified by domain experts using our approach.}
    \label{fig:heatmap+daria+ours}
\end{figure}

\begin{figure}
    \centering
    \includegraphics[width=\linewidth]{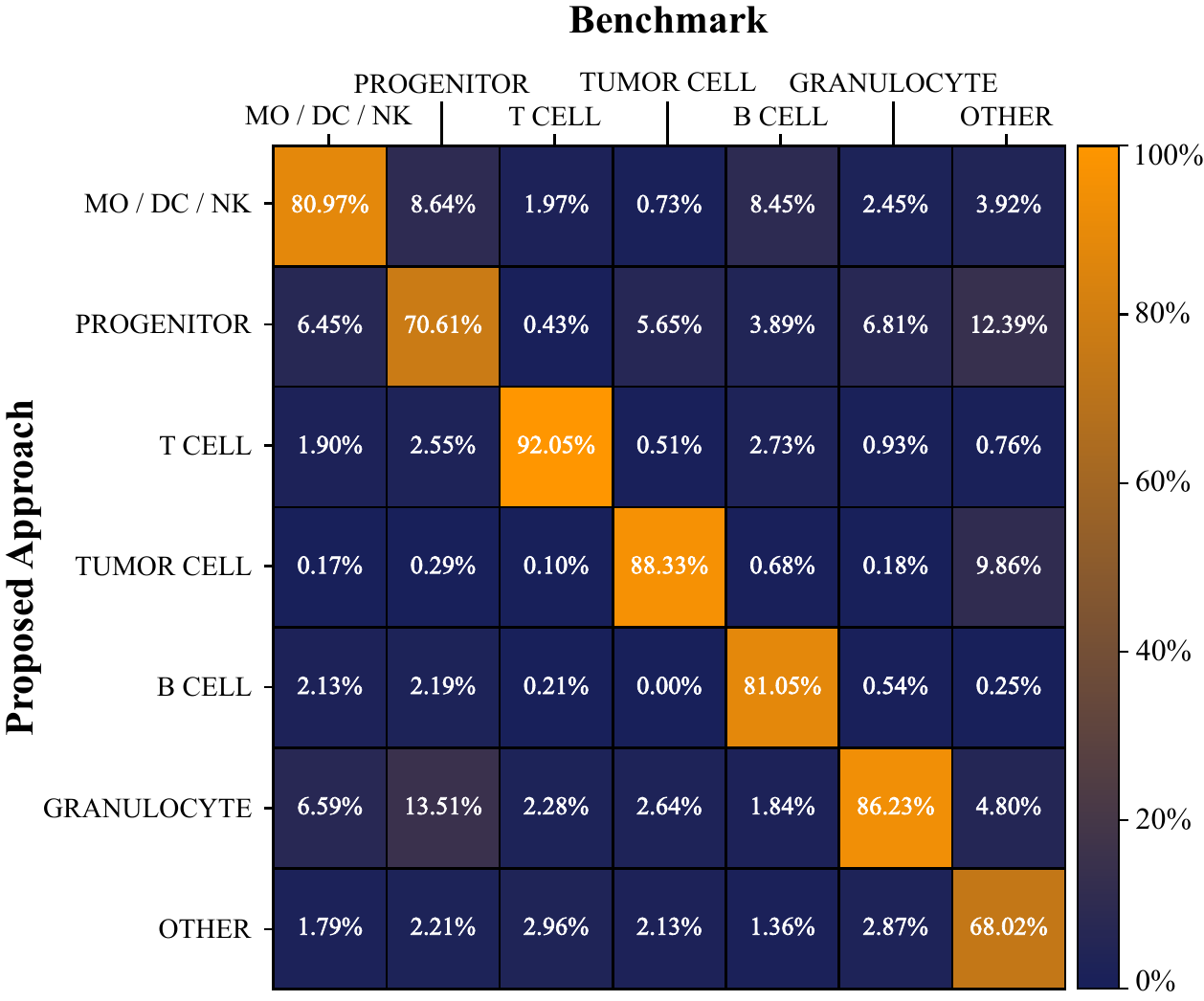}  
    \caption{Rediscovery rate of each cell type cluster identified in the benchmark analysis using segmentation, compared to the cell type clusters assigned by domain experts using our proposed method. Normalized by row.}
    \label{fig:confusion_matrix}
\end{figure}

\noindent Following the qualitative assessment of the UMAPs, the quantitative analysis of cell type rediscovery rates confirms that biologically relevant cells effectively aggregate in the representation space using the proposed method. Phenograph clustering produced 40 clusters, each assigned to one of the predefined cell type clusters. Fig.~\ref{fig:heatmap+daria+ours}~B shows the annotations from the benchmark analysis using segmentation and manual feature extraction, while Fig.~\ref{fig:heatmap+daria+ours}~C presents the cluster assignments obtained using our proposed approach. There is a strong agreement between the two, with accurate identification of T cells, B cells, tumor cells, and granulocytes. This is validated by the confusion matrix (Fig.~\ref{fig:confusion_matrix}), which shows high rediscovery rates for T cells (92.05\%), B cells (81.05\%), tumor cells (88.33\%), and granulocytes (86.23\%).\\
Classifying progenitor cells is particularly challenging due to their transitional state and overlapping marker expression with both precursor and mature cell types. This overlap causes discrepancies, with 13.51\% of progenitors being classified as granulocytes and 8.64\% as MO/DC/NK cells. These discrepancies underscore the complexity of progenitor cells, which exist along a differentiation continuum and are not fully definable by the markers included in the IMC panel. This also raises questions about the accuracy of progenitor cell annotations in the benchmark analysis.\\

\begin{figure}
    \centering
    \includegraphics[width=\linewidth]{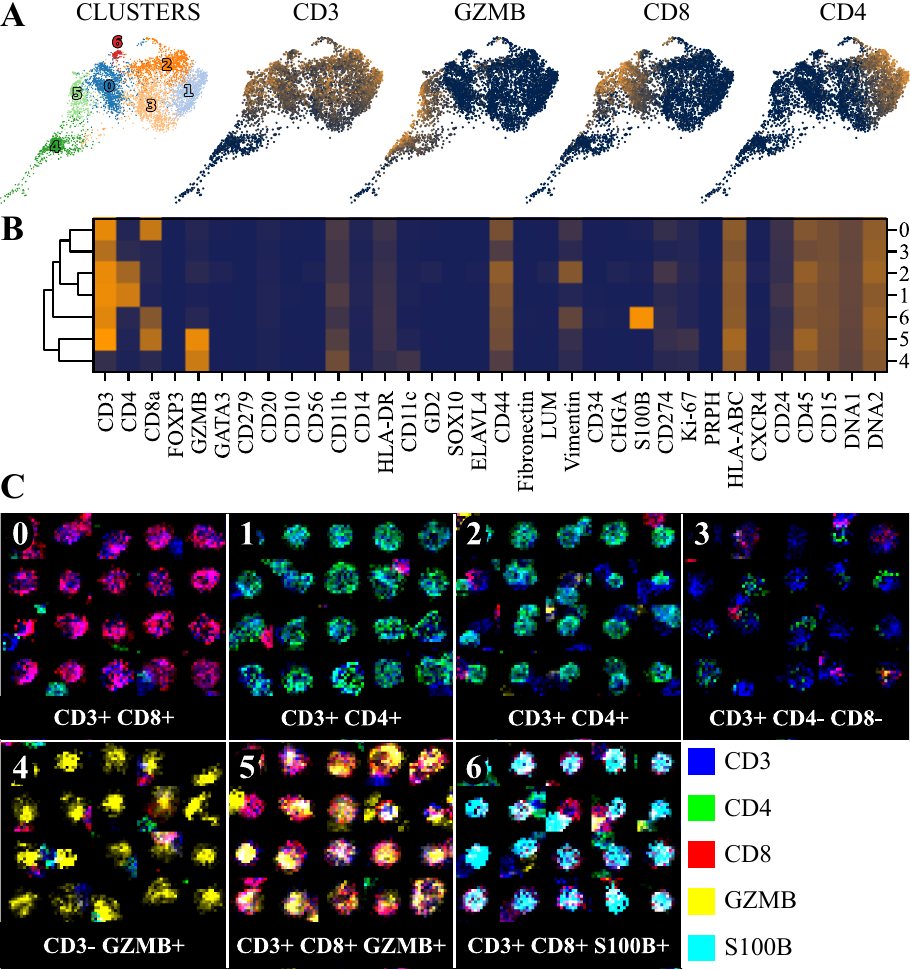}  
    \caption{\textbf{A}: Detailed UMAP visualization highlighting distinct T cell subclusters identified by Phenograph, showing the activation obtained from the interpretability stage of our proposed approach for key T cell markers: CD3, GZMB, CD8, and CD4. \textbf{B}: Average activation per subcluster for all 34 markers in the panel. \textbf{C}: Gallery of representative cells from each of the 7 subclusters, with markers displayed in distinct colors.}
    \label{fig:t-cells}
\end{figure}

\noindent\textbf{Subdivision of T-cell cluster}: We further demonstrate the interpretability and utility of our approach by subclustering the T-cell cluster, composed of a set of subtypes with distinct signatures  (Fig.~\ref{fig:heatmap+daria+ours}~A). Within the T cell cluster, we identified seven distinct subclusters (Fig.~\ref{fig:t-cells}~A), each corresponding to well-characterized cell types reported in the literature. These include: (0) CD3+ CD8+ cytotoxic T cells and (1) CD3+ CD4+ Vimentin- helper T cells, (2) CD3+ CD4+ Vim+ activated/memory T cells, (3) CD3+ CD8- CD4- double negative T cells, (4) CD3- GZMB+ natural killer cells, and (5) CD3+ CD8+ GZMB+ cytotoxic T cells and (6) CD3+ CD8+ S100B+ cytotoxic T cells,  where “+” denotes marker expression and “-” indicates its absence. To validate that these clusters indeed represent cells with the expression patterns shown in the heatmap (Fig.~\ref{fig:t-cells}~B), we present a gallery of patches from representative cells within each subcluster (Fig.~\ref{fig:t-cells}~C).\\
This granularity in identifying T-cell subpopulations showcases the model’s ability to resolve biologically meaningful subtypes beyond broad cell type distinctions. The clear separation of these subpopulations underscores the quality and interpretability of the representation space, demonstrating its suitability for detailed phenotypic characterization. %
\section{Conclusion} We introduced a segmentation-free deep learning approach that leverages grouped convolutions to learn biologically interpretable features from IMC data, overcoming the limitations of segmentation in low-resolution images and eliminating the need for extensive fine-tuning. Validated on an IMC dataset of 1.8 million bone marrow cells, our method successfully rediscovered known cell types and distinguished biologically meaningful subpopulations, including T-cell subtypes. This scalable and interpretable framework offers a powerful solution for high-dimensional single-cell analysis, with potential applications across diverse imaging modalities.
%
%
%
%






%
\clearpage
\section{Ethics Statement}
Ethical approval for the use of bone marrow aspirates from the CCRI Biobank, and clinical data was granted by the local institutional review board (IRB) of the Medical University of Vienna (EK1224/2020).

\bibliographystyle{IEEEbib}
\bibliography{main.bib}

\end{document}